\documentclass[%
superscriptaddress,
twocolumn,
 amsmath,amssymb,
 aps,
prb,
]{revtex4-1}

\usepackage{graphicx}
\usepackage{bm}
\usepackage{hyperref}

\renewcommand{\v}[1]{\ensuremath{\mathbf{#1}}} 
\newcommand{\gv}[1]{\ensuremath{\mbox{\boldmath$ #1 $}}}
\newcommand{\pd}[2]{\frac{\partial #1}{\partial #2}}

\newcommand{\grad}[1]{\gv{\nabla} #1} 
\renewcommand{\div}[1]{\gv{\nabla} \cdot #1} 
\newcommand{\curl}[1]{\gv{\nabla} \times #1} 
\let\baraccent=\= 
\renewcommand{\=}[1]{\stackrel{#1}{=}} 

\newcommand{\nn}{\nonumber \\}

\begin{document}

\title{Nonlocal transport phenomena in Weyl metals beyond the mesoscopic scale}

\author{Jinho Yang}

\affiliation{Department of Physics, POSTECH, Pohang, Gyeongbuk 37673, Korea}

\author{Ki-Seok Kim}

\affiliation{Department of Physics, POSTECH, Pohang, Gyeongbuk 37673, Korea}

\affiliation{Asia Pacific Center for Theoretical Physics (APCTP), Pohang, Gyeongbuk 37673, Korea}

\date{\today}

\begin{abstract}
Axion electrodynamics governs electromagnetic properties of Weyl metals. Although transmission and reflection measurements of light have been proposed to confirm the axion electrodynamics, there are still lack of theoretical proposals for macroscopic nonlocal transport phenomena in Weyl metals. In this paper, we present nonlocal transport phenomena in time reversal symmetry-broken (TRSB) Weyl metals. Solving the axion electrodynamics numerically, we show that such nonlocal transport phenomena arise from the negative longitudinal magneto-resistivity (NLMR), combined with the anomalous Hall effect (AHE) in the axion electrodynamics. Since this nonlocal transport occurs beyond the mesoscopic scale, we conclude that these nonlocal properties have nothing to do with Fermi arcs, regarded to be clear evidence of the axion electrodynamics in the bulk.
%
%
\end{abstract}

\maketitle

\section{Introduction}

Maxwell equations are modified in Weyl metals \cite{WM1,WM2,WM3,WM4}, which originate from anomalous electromagnetic currents \cite{CME1,CME2,CME3,CME4,CME5,CME6,CME7,Boltzmann_Chiral_Anomaly1,Boltzmann_Chiral_Anomaly2,Boltzmann_Chiral_Anomaly3,Boltzmann_Chiral_Anomaly4,
Boltzmann_Chiral_Anomaly5,Boltzmann_Chiral_Anomaly6,Boltzmann_Chiral_Anomaly7,Boltzmann_Chiral_Anomaly8,Boltzmann_Chiral_Anomaly9,AHE1,AHE2,AHE3,AHE4}. More concretely, the topological-in-origin $\v{E} \cdot \v{B}$ term with a spacetime dependent coefficient $\theta(\bm{r},t)$ angle occurs in the effective action for electromagnetic fields from the so called chiral anomaly, which gives rise to corrections in the Maxwell equation \cite{Axion_EM}. Although axions as dynamical degrees of freedom have been proposed in various situations such as charge and spin density wave orders, ferromagnetism, and superconductivity \cite{Dynamical_Axion_Review}, we focus on the case of external non-dynamical fields for axions in this study.

There exist theoretical proposals to confirm the axion electrodynamics. In particular, transmission and reflection experiments of light have been proposed to measure Faraday and Kerr rotations or higher harmonics in Weyl metals \cite{Axion_EM_Exp_TI_I,Axion_EM_Exp_TI_II,Axion_EM_Exp_TI_III,AxionEMreferee1,AxionEMreferee2,AxionEMreferee3,Axion_EM_Th_WM,AxionEMreferee4}. Non-linear effects have been mainly focused on the second harmonic generation in the optical regime \cite{nonlinear1,nonlinear2,nonlinear3,nonlinear4} (i.e., the frequency of an oscillating field is in $\omega > 20$ kHz). According to one theoretical proposal, the axion electrodynamics allows a longitudinal component inside the Weyl metallic state as superconductivity does \cite{Axionprb}. However, we believe that there are still lack of theoretical proposals for macroscopic nonlocal transport phenomena in Weyl metals. In this paper, we present nonlocal transport phenomena in time reversal symmetry-broken (TRSB) Weyl metals.

Nonlocal transport properties in TRSB Weyl metals have been reported before \cite{nonlocaldiffuse,nonlocalPW,nonlocaleeinter,nonlocalhydro1,nonlocalhydro2}. However, to our best knowledge, such nonlocal effects are limited in the mesoscopic scale related to the diffusive origin \cite{nonlocaldiffuse}, or they occur from the combination of the topological origin and others (e.g., artificial potential wall \cite{nonlocalPW} or electron-electron interaction \cite{nonlocaleeinter,nonlocalhydro1,nonlocalhydro2}). In this study, we show that nonlocal transport phenomena are allowed in the macroscopic level within the axion electrodynamics.
%
%
Solving the axion electrodynamics numerically, we reveal that macroscopic nonlocal voltage drop is possible due to geometrically asymmetric conductivity in Weyl metals. Such asymmetry in conductivity turns out to result from the negative longitudinal magneto-resistivity (NLMR) \cite{TSB_WM1,TSB_WM2,ISB_WM1,ISB_WM2,ISB_WM3,ISB_WM4,ISB_WM5,ISB_WM6,ISB_WM7} in combination with the transverse magneto-resistivity (TMR) \cite{TMR} and the anomalous Hall effect (AHE) \cite{CME1,CME2,CME3,CME4,CME5,CME6,CME7,Boltzmann_Chiral_Anomaly1,Boltzmann_Chiral_Anomaly2,Boltzmann_Chiral_Anomaly3,Boltzmann_Chiral_Anomaly4,
Boltzmann_Chiral_Anomaly5,Boltzmann_Chiral_Anomaly6,Boltzmann_Chiral_Anomaly7,Boltzmann_Chiral_Anomaly8,Boltzmann_Chiral_Anomaly9,AHE1,AHE2,AHE3,AHE4}. When external magnetic fields $\v{B_{ext}}$ are applied, electric currents parallel to $\v{B_{ext}}$ are enhanced by the $B_{ext}^2$ factor (NLMR). On the other hand, currents perpendicular to $\v{B_{ext}}$ is proportional to $1/B_{ext}$ (TMR). Furthermore, there exists the AHE for the direction perpendicular to the external magnetic field. The asymmetric conductivity induced by these effects results in nonlocal and non-homogeneous electric fields/currents.

To simulate these effects in the TRSB Weyl metal state, we incorporate both the NLMR and TMR into the conductivity $\sigma$ of the Ohm's law $\v{J} = \sigma \v{E}$. In addition, we consider the conserved current $\v{J}' = \v{J} - 2 \alpha \grad \theta \times \v{E}$ to take into account the current from the Fermi surface ($\v{J}$) with both magneto-resistivity contributions and that from the AHE ($- 2 \alpha \grad \theta \times \v{E}$). Here, $\alpha$ is the fine-structure constant.
%
%

\section{Simulation setup for the axion electrodynamics}

We explain how to simulate the axion electrodynamics, introducing the Maxwell equation with a modified conserved current $\v{J}'$ in a discrete grid. We point out that a simulation procedure for 2D conventional metals is shown in appendix \ref{2Dcase} as a pedagogical example. Essentially the same strategy is applied to the case of 3D Weyl metals, presented in the last of this section.

\subsection{Axion electrodynamics}

\begin{figure}
\centering
\includegraphics[width=8cm]{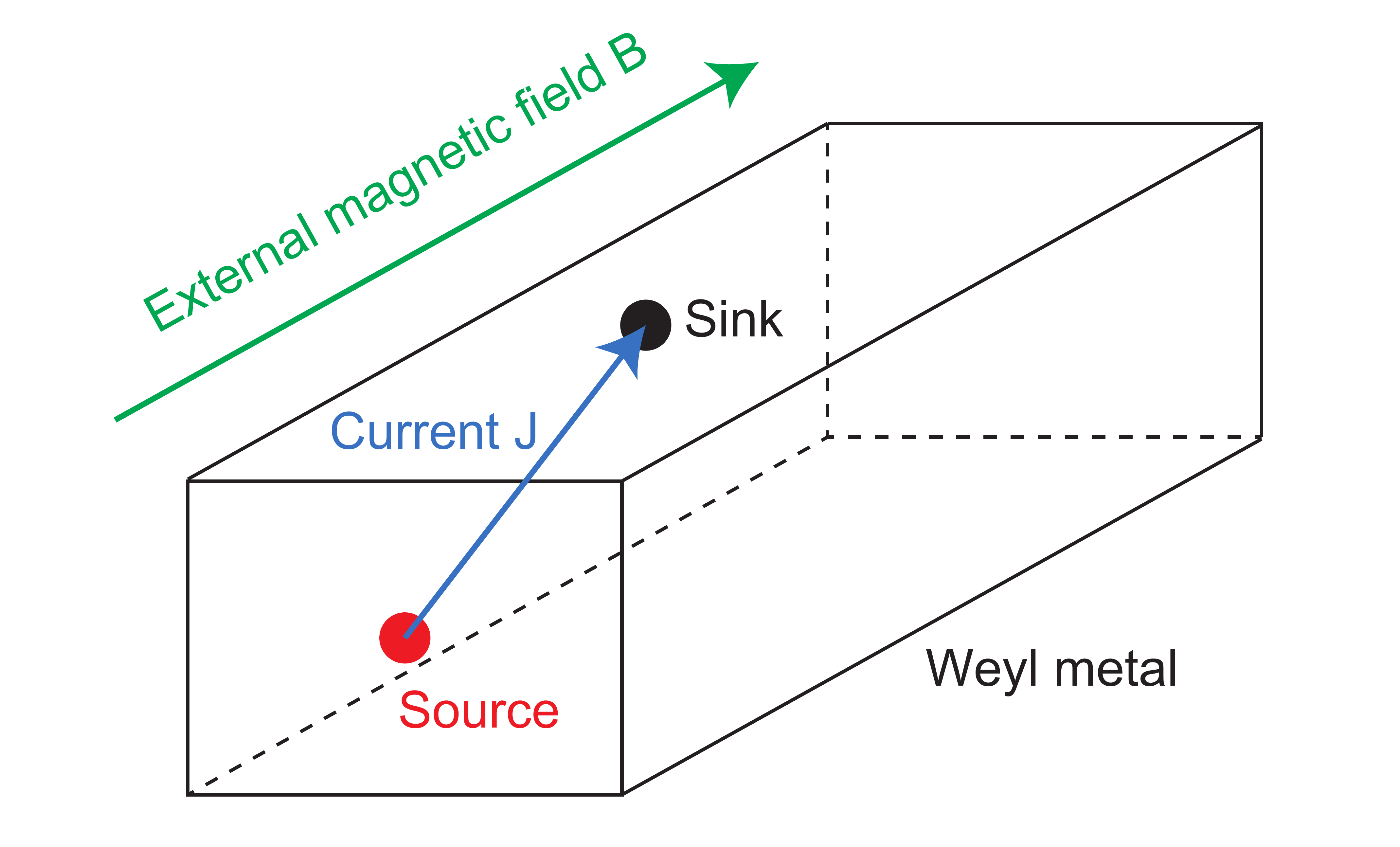}
\caption{Schematic diagram for the simulation of nonlocal electric-field measurements based on the axion electrodynamics. A direct current path is shown by the blue arrow, which starts from a source point (red dot) to a sink point (black dot). Our simulations on the axion electrodynamics under an external magnetic field (green arrow) reveal that there appear inhomogeneously distributed nonlocal currents in macroscopic-size 3D Weyl metals.}
\label{setting}
\end{figure}

We start from the axion electrodynamics in an experimental setup of Fig. \ref{setting},
\begin{eqnarray}
\div \textbf{E} &=& \rho/\epsilon + 2\alpha/\epsilon \grad \theta \cdot \textbf{B} , \label{Maxwell1} \\
\div \textbf{B} &=& 0 , \label{Maxwell2} \\
\curl \v{E} &=& -\pd{\v{B}}{t} , \label{Maxwell3} \\
\curl \v{B} &=& \mu \v{J} + \mu \epsilon \pd{\v{E}}{t} - 2\alpha \mu \grad{\theta} \times \v{E} .
\label{Maxwell4}
\end{eqnarray}
As shown in this experimental setup, a direct current path is given by the blue arrow, which starts from a source point (red dot) to a sink point (black dot). On the other hand, our simulations on the axion electrodynamics under an external magnetic field (green arrow) reveal that there appear inhomogeneously distributed nonlocal currents in macroscopic-size 3D Weyl metals. This originates from contributions of anomalous currents in the axion electrodynamics.

When $\grad \theta$ is proportional to uniformly applied magnetic fields ($\v{B_{ext}}$) \cite{CME1,CME2,CME3,CME4,CME5,CME6,CME7,Boltzmann_Chiral_Anomaly1,Boltzmann_Chiral_Anomaly2,Boltzmann_Chiral_Anomaly3,Boltzmann_Chiral_Anomaly4,
Boltzmann_Chiral_Anomaly5,Boltzmann_Chiral_Anomaly6,Boltzmann_Chiral_Anomaly7,Boltzmann_Chiral_Anomaly8,Boltzmann_Chiral_Anomaly9,AHE1,AHE2,AHE3,AHE4}, meaning constant in space and time, we obtain
\begin{widetext}
\begin{eqnarray}
 \div{(\textrm{\ref{Maxwell4}})} &=& \div{(\curl{\v{B}})} =  0 \rightarrow \mu \div{\v{J}} + \mu \epsilon \pd{\div \v{E}}{t} - 2\alpha \mu \div \v{\grad{\theta}} \times \v{E} = 0. \nn
\pd{\epsilon \div \v{E}}{t} &=& - \div{\v{J}} + 2 \alpha \div{(\grad{\theta} \times \v{E})} \nn
&=&  - \div{\v{J}} +2\alpha \mu (\v{E} \cdot (\curl{\grad \theta}) - \v{\grad \theta} \cdot (\curl{\v{E}})) \nn
&=&  - \div{\v{J}} +2\alpha (\v{\grad \theta} \cdot \pd{\v{B}}{t}), \nonumber \\
\nonumber \\
\therefore \div \v{E} &=& - \int \frac{ \div{\v{J}} + 2\alpha \grad{\theta} \cdot \pd{\v{B}}{t}}{\epsilon}dt = \rho/ \epsilon + 2\alpha/ \epsilon \grad{\theta} \cdot \v{B} + C_0 (\v{x,y,z}). \\
 \div{(\textrm{\ref{Maxwell3}})} &=& \div{(\curl{\v{E}})} = \pd{\div{\v{B}}}{t} = 0, \nn
\therefore \div{\v{B}} &=& \int \pd{\div{\v{B}}}{t} dt = C_1(\v{x,y,z}) .
\end{eqnarray}
\end{widetext}
Here, divergence has been taken for Eqs. (\ref{Maxwell3}) and (\ref{Maxwell4}), and $\div{(\v{A} \times \v{B})}=\v{B} \cdot (\curl{\v{A}}) - \v{A}\cdot \curl{\v{B}}$ has been used. Since the first and second Maxwell equations should be consistent with the third and fourth Maxwell equations, we conclude that both constants of $C_0$ and $C_1$ have to vanish identically, i.e., $C_0$ = $C_1$ = 0.

For more general discussions, let us replace $\grad{\theta}$ with $g\v{B}(t)$ in the axion electrodynamics, time dependent magnetic fields. Then, the first Maxwell equation is given from the fourth Maxwell equation (Eq. (\ref{Maxwell4})) as follows
\begin{widetext}
\begin{eqnarray}
\div{(4)} &=& \div{(\curl{\v{B}})} =  0 \rightarrow \div{(\mu \v{J} + \mu \epsilon \pd{\v{E}}{t} - 2\alpha g \mu \v{{\v{B}}} \times \v{E})} = 0\nn
\pd{\div{\epsilon \v{E}}}{t} &=& -\div{\v{J}} + 2 \alpha g \div{(\v{B} \times \v{E})} \\
&=& -\div{\v{J}} - 2\alpha g \mu(\div{\v{S}}) \\
\nonumber \\
\therefore \div{\epsilon \v{E}} &=& - \int \div{(\v{J} + 2\alpha g \mu \v{S} )}dt .
\end{eqnarray}
\end{widetext}
Here, the constant $C_0$ from the time integral is set to be zero, as discussed above. We point out that the Poynting theorem \cite{EM_Textbook} given by $-\partial_t u_{em} = \div{\v{S}} + \v{J} \cdot \v{E}$ is still satisfied in the axion electrodynamics, as shown in appendix \ref{Poynting}. Here, $\v{S} \equiv \frac{1}{\mu} \v{E} \times \v{B}$ is the Poynting vector and $u_{em} = \frac{\v{B}^2}{2\mu} + \frac{\epsilon \v{E}^2}{2}$ is the electromagnetic field energy.

Using the Poynting vector $\v{S} \equiv \frac{1}{\mu} \v{E} \times \v{B}$, we generalize the above axion electrodynamics as follows
\begin{eqnarray}
\div \v{E} &=&  - \frac{1}{\epsilon}\int \div{(\v{J} + 2\alpha g \mu \v{S})} dt \label{newMaxwell1} \\
\div \v{B} &=& 0 \label{newMaxwell2}\\
\curl \v{E} &=& -\pd{\v{B}}{t} \label{newMaxwell3}\\
\curl \v{B} &=&  \mu \v{J} + \mu \epsilon \pd{\v{E}}{t} + 2\alpha g \mu^2 \v{S}.  \label{newMaxwell4}
\end{eqnarray}

Interestingly, these equations can be rewritten in the form of the original Maxwell equations
\begin{eqnarray}
\div \v{E} &=& -\frac{1}{\epsilon} \int \div \v{J'} dt = \frac{\rho'}{\epsilon} \\
\div \v{B} &=& 0 \\
\curl \v{E} &=& -\pd{\v{B}}{t} \\
\curl \v{B} &=& \mu \v{J'} + \mu \epsilon \pd{\v{E}}{t} ,
\end{eqnarray}
introducing anomalous current density $\v{J'}$ and charge density $\rho'$ in the following way
\begin{eqnarray}
\rho' &\equiv& - \int \div{\v{J'}} dt = - \int \div{(\v{J} + 2\alpha g \mu \v{S})} dt \nonumber \\
         &=& - \int \div{\v{J}} dt + 2\alpha g \mu (u_{em} + W) , \label{totalrho} \\
\v{J'} &\equiv& \v{J} + 2\alpha g \v{E} \times \v{B} . \label{conservedJ}
\end{eqnarray}
Here, $W = \int \v{E} \cdot \v{J} dt$ represents work done by the current source. These equations show that the anomalous current and charge take into account the angular momentum and energy density from electromagnetic fields.

Of course, these anomalous charge density $\rho'$ and current density $\v{J}'$ satisfy the continuity equation
\begin{eqnarray}
\div \v{J'} &=& -\pd{\rho'}{t} \nonumber \\
                  &=& \div{\v{J}} - 2 \alpha g \mu \left \{ \partial_t \left (\frac{\v{B}^2}{2\mu} + \frac{\epsilon \v{E}^2}{2} \right) + \v{E} \cdot \v{J} \right \} . \nonumber \\ \label{continuity}
\end{eqnarray}
We note that $\div \v{J}$ is not zero but $\div \v{J'}$ should be zero considering time dependence in $\v{E}$ and $\v{B}$. This indicates that $\v{J}'$ should be regarded as the conserved current satisfying the continuity equation instead of $\v{J}$.
%
%
In this study, we use $\v{J'}$ as the conserved current for all simulations.

\subsection{Simulation setup for the axion electrodynamics}

\begin{figure*}
\centering
\includegraphics[width=16cm]{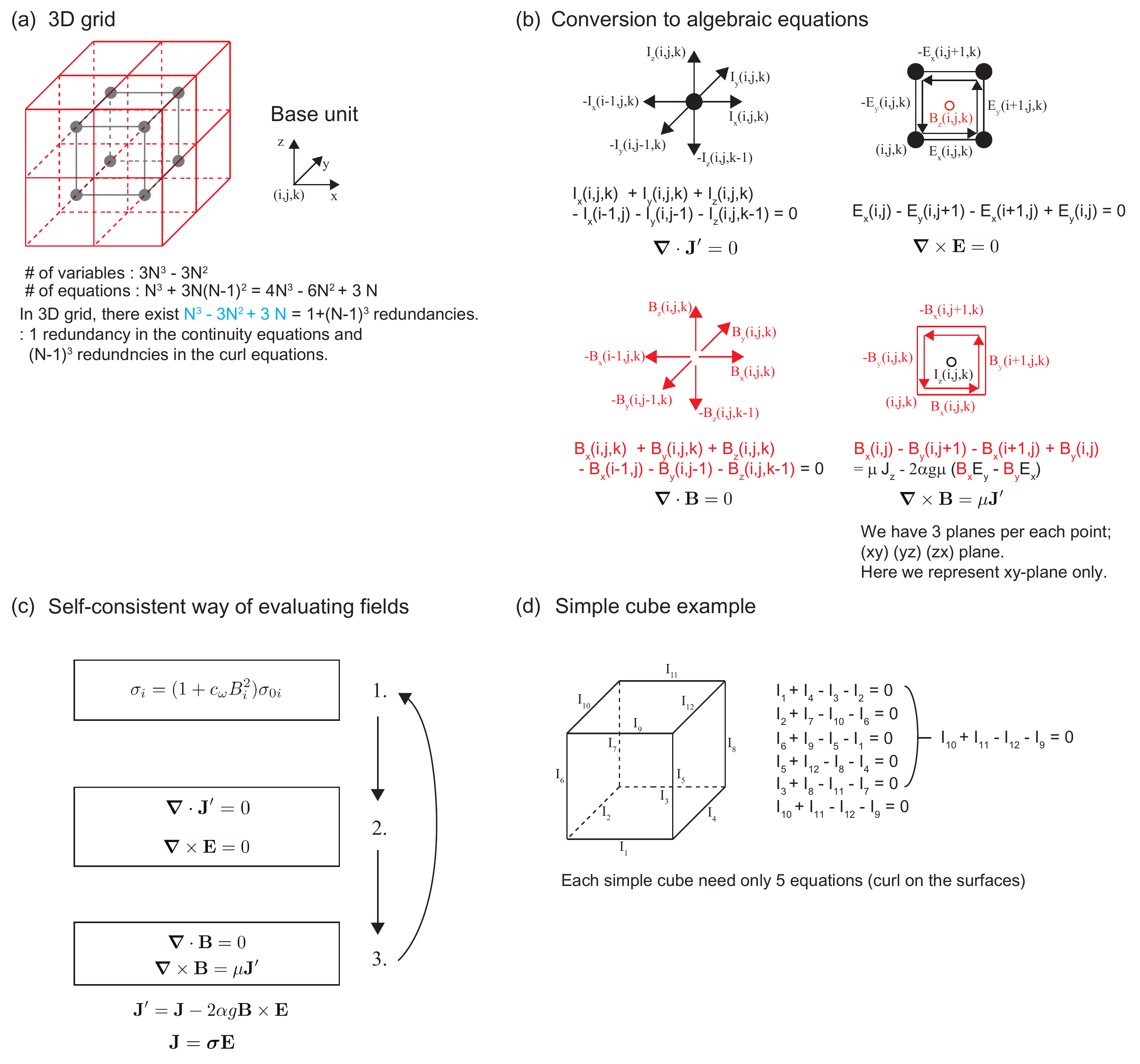}
\caption{How to simulate the axion electrodynamics in 3D grid. (a) 3D grid structure with electromagnetic-field and current variables. Here, the current and conductivity are defined on black links whereas the magnetic field is defined on red links with unit vectors $\hat{x}$, $\hat{y}$, $\hat{z}$. Number of variables, equations, and constraints are shown. See the text for more details. (b) Conversion of divergence and curl equations into algebraic equations on the grid structure. Divergence is converted as sum of six vectors at every vertex, whereas curl is converted as sum of four vectors at every face of the unit cube. (c) Simulation protocol for anomalous currents $I$, magnetic fields $B$, and magnetoconductivity $\sigma$. For more details, see the text. (d) A simple example of redundancy in curl equations (unit cube with 6 surfaces). There is one redundancy per one unit cube. There are 6 curl equations from 6 planes for one unit cube, but summing over 5 equations gives the other one.}
\label{3dsim}
\end{figure*}

To investigate the nonlocal transport phenomena in a Weyl metal, we recall that both longitudinal and transverse magnetoresistivities and anomalous Hall effect have to be taken into account in the fourth Maxwell equation ($\curl \v{B} = \mu \v{J} - 2 \alpha g \mu \v{B} \times \v{E}$). To find the electric field $\v{E}$ and the magnetic field $\v{B}$ with this anomalous current in a numerical way, we define such electromagnetic fields and anomalous currents on vertices, edges, and faces of a 3D grid structure (cubic lattice). The conductivity $\boldsymbol{\sigma}(i,j,k)$ and current $\v{I}(i,j,k)$ are defined on black links whereas the magnetic field $\v{B}(i,j,k)$ is defined on red links of the 3D grid. See Fig. \ref{3dsim} (a). Now, the black and red grids are conjugate (dual) to each other. The divergence (Eqs. (\ref{Maxwell1}) and (\ref{Maxwell2})) and curl equations (Eqs. (\ref{Maxwell3}) and (\ref{Maxwell4})) in the Maxwell equations can be converted into algebraic forms as shown in Fig. \ref{3dsim} (b). One may think that the current (magnetic field) is defined on each lattice point of the body center of the red (black) links for three directions. For more detailed information on this construction, we refer it to appendix \ref{2Dcase}, the construction of which is discussed in a simpler situation on 2D grid.

Now, we solve coupled algebraic equations for electromagnetic-field variables with anomalous currents, defined on each unit cube (and its dual) in the grid.
%
%
The iteration process consists of three steps. First, the longitudinal magnetoconductivity is considered with the Ohm's law for the current of chiral Fermi-surface electrons. Here, an initial value ($\v{B} = \v{B_{ext}}$) of the magnetic field is used for this conductivity. Second, the divergence/curl of the electric field is evaluated with a given conductivity of the first step. Third, the divergence/curl of the magnetic field is evaluated using the current given at the second step. We note that the current is determined once both conductivity and electric field are given by the Ohm's law with the anomalous Hall effect. Updating the magnetic field in the longitudinal magnetoconductivity of the first step, we repeat this process for convergence. See Fig. \ref{3dsim} (c).

Before presenting our simulation results, we would like to mention two important aspects in 3D Weyl-metal simulations different from the 2D simulation (see appendix \ref{2Dcase} for the 2D case). First, one is on the redundancy in the number of equations simply due to the additional component of $z$. Considering divergence equations with continuity equations on $N^3$ points (vertices) and curl equations on $(N-1)^3$ planes (faces), all these equations cannot be independent. The continuity equations have one redundancy the same as that of the 2D case. On the other hand, we have $(N-1)^3$ redundancies in curl equations for the 3D case (no redundancy in curl equations for the 2D case). Here, we explain how to count the number of redundancies in the curl equations. Each single unit cube allows only five curl equations (5 independent surfaces) to define all variables. See Fig. \ref{3dsim} (d). For simplicity, let us assume without loss of generality that redundant equations are living in the $xy$ plane of the upper layer for each single cube. It means that the curl equations in $z=1$, $z=2$, ..., $z=N-1$ (layer) are all redundant for the $N \times N \times N$ lattice ($x$, $y$, and $z$ are integer values from $0$ to $N-1$). Therefore, the curl equations on $(N-1)$ layers are redundant. Each layer has $(N-1)^2$ curl equations, and thus, we have $(N-1)^3$ redundancies in the curl equations. In total, we have $1 + (N-1)^3 $ redundancies in the number of equations for the 3D case.

The second important point in the 3D case is on anomalous currents originating from the $\grad \theta$ (external or background axion) term. In 2D, the Maxwell equation to govern the dynamics of electric fields is given by the Ohm's law ($\v{E} = \boldsymbol{\sigma}^{-1} \v{J}$). On the other hand, one should consider $\v{J'}$ with the AHE as the conserved current, given by
\begin{eqnarray}
\v{J'} = \v{J} - 2 \alpha g \v{B} \times \v{E} , \nonumber
\end{eqnarray}
as discussed before. Considering the linear expansion for the electric field ($\v{J} = \boldsymbol{\sigma} \v{E}$), we express $\v{J}$ in terms of $\v{J'}$ as follows
\begin{eqnarray}
J_x &=& J'_x + 2 \alpha g (B_y J_z/\sigma_z - B_z J_y/\sigma_y) \nonumber \\
J_y &=& J'_y + 2 \alpha g (B_z J_x/\sigma_x - B_x J_z/\sigma_z) \nonumber \\
J_z &=& J'_z + 2 \alpha g (B_x J_y/\sigma_y - B_y J_x/\sigma_x). \label{selfconsistentJ}
\end{eqnarray}
Here, we ignored the conventional Hall conductivity. As a result, the electric field $\v{E}$ is given by
\begin{eqnarray}
E_x = \frac{J'_x}{\sigma_x} + \frac{2 \alpha g}{\sigma_x} \left \{ B_y \frac{J'_z}{\sigma_z} - B_z \frac{J'_y}{\sigma_y}\right \} + (2 \alpha g)^2 \{ \cdot \cdot \cdot \} + \cdot \cdot \cdot \nonumber \\
E_y = \frac{J'_y}{\sigma_y} + \frac{2 \alpha g}{\sigma_y} \left \{ B_z \frac{J'_x}{\sigma_x} - B_x \frac{J'_z}{\sigma_z}\right \} + (2 \alpha g)^2 \{ \cdot \cdot \cdot \} + \cdot \cdot \cdot \nonumber \\
E_z = \frac{J'_z}{\sigma_z } + \frac{2 \alpha g}{\sigma_z } \left \{ B_x \frac{J'_y}{\sigma_y } - B_y \frac{J'_x}{\sigma_x }\right \} + (2 \alpha g)^2 \{ \cdot \cdot \cdot \} + \cdot \cdot \cdot . \nonumber \\
 \label{selfconsistentE}
\end{eqnarray}
%
%
If we set $2 \alpha g = 0$, this expression is reduced to the current with only magnetoresistivity in the absence of the AHE. In the next section, we discuss the nonlocal transport phenomena in both cases of the presence and absence of the AHE.

\section{nonlocal transport phenomena from asymmetric conductivity in the axion electrodynamics}

\subsection{In the absence of the anomalous Hall effect}

\begin{figure*}
\centering
\includegraphics[width=15cm]{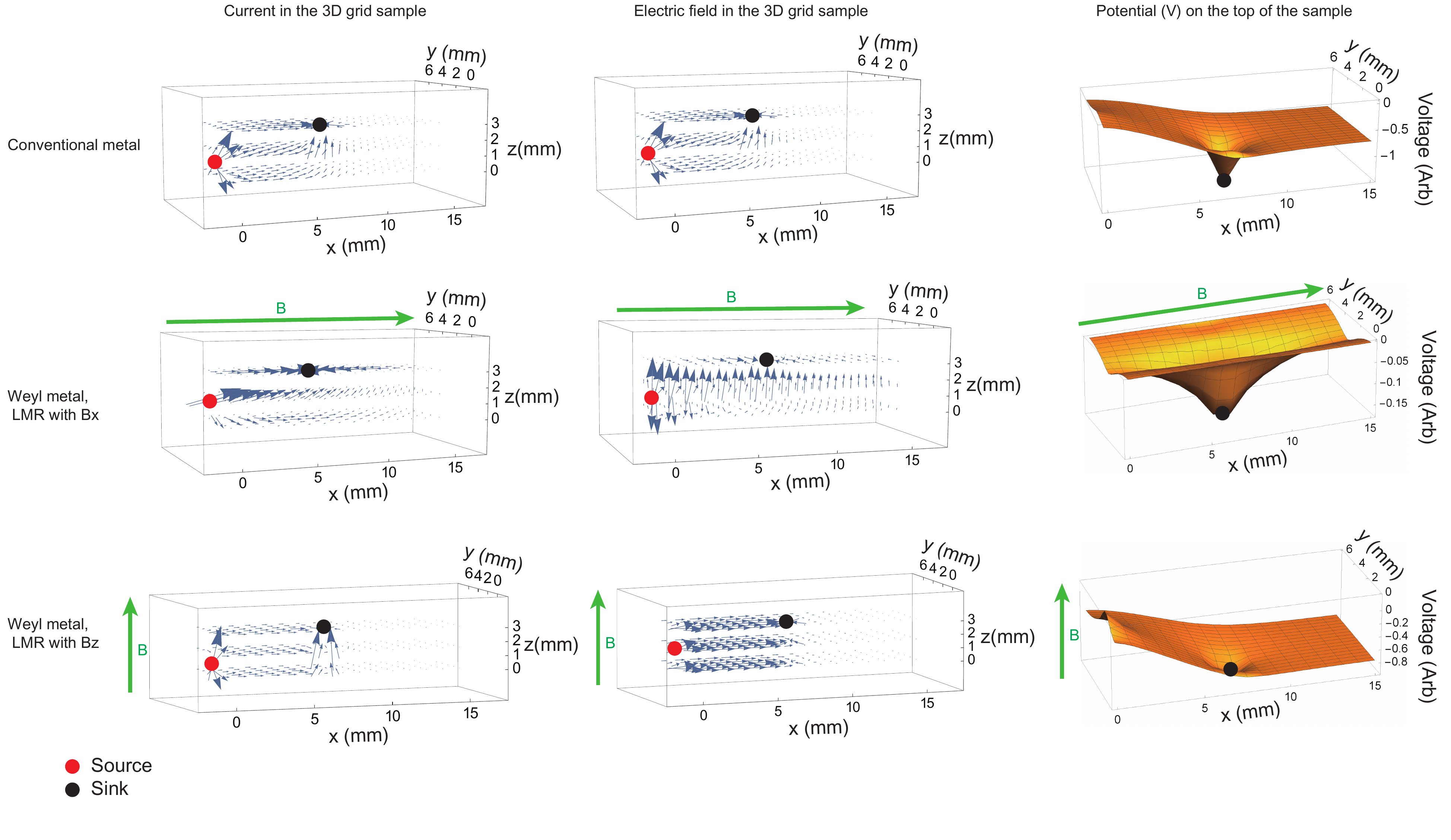}
\caption{``Partial" axion electrodynamics simulation in the absence of the anomalous Hall effect. The first (second) column shows the current (electric field) configuration, and the last describes a potential landscape. In isotropic conventional metals shown in the first row, nonlocal currents are not observed. On the other hand, the NLMR effect in a Weyl metal state, given in the second row, is responsible for the nonlocal voltage drop in a $mm$ scale of the sample size. In the transverse configuration given in the last row, nonlocal currents are not observed, either. See the text for more details.}
\label{3dgradthetatotal}
\end{figure*}

First, we take into account only the magnetoresistivity without the AHE, i.e., ignoring the $\grad \theta \times \v{E}$ term. There are two types of magnetoresistivity effects. One is NLMR given by $\sigma_{\parallel \v{B}} = \sigma_0 (1 + c \v{B_{ext}}^2)$, which shows the $B^2$ enhancement for the longitudinal conductivity. Here, $\sigma_0$ is the Drude conductivity and $c$ is a dimensionful constant. The other is TMR given by $\sigma_{\perp \v{B}} \propto 1/B_{ext} $, the magnetoconductivity of which is proportional to inverse of the magnetic field in the large magnetic-field limit. The transverse magnetoconductivity is much smaller than the longitudinal one in the limit of large magnetic fields. These geometrically different two magnetoconductivities can change both the current and electric-field configurations dramatically, as shown in Fig. \ref{3dgradthetatotal}. Quantitatively speaking, the external magnetic field is set to make the conductivity ratio as $\sigma_{\parallel \v{B}} : \sigma_{\perp \v{B}} = 1 : 20$ in this simulation.

In isotropic conventional metals, both current and electric-field configurations are essentially the same as each other. In other words, their dominant flows are given by a straight and direct line configuration from the source to the sink. On the other hand, when an external magnetic field is applied in parallel with the source to the sink direction, i.e., $\v{B_{ext}} \parallel \v{J_{ext}}$ (longitudinal) in a Weyl metal, the current flow becomes nonlocal in the $x$ direction. Here, ``nonlocal" means that the current flow or voltage drop exists not only between the source and sink but also far from the source and sink positions. Electric field becomes predominant along the $y$ direction due to the NLMR effect because the current flow requires only weak strength of the electric field. On the other hand, the current flow gives rise to relatively stronger electric fields when the current flows orthogonal to the external magnetic-field direction. In this transverse case, we don't see any nonlocal effects. All these results are summarized in Fig. \ref{3dgradthetatotal}.

It is not easy to construct either perfectly parallel or absolutely perpendicular setup for any configurations of external currents and magnetic fields in a real experimental condition. In this respect it is interesting to ask whether this nonlocal current can be observed or not in slightly tilted configurations as our simulation setting. It turns out that such small tilted angle configuration does not give any significant effects in this case. However, we find that it gives nonlocal Hall effects when the AHE is taken into account, discussed in the next subsection.

\subsection{In the presence of the anomalous Hall effect}

\begin{figure*}
\centering
\includegraphics[width=16cm]{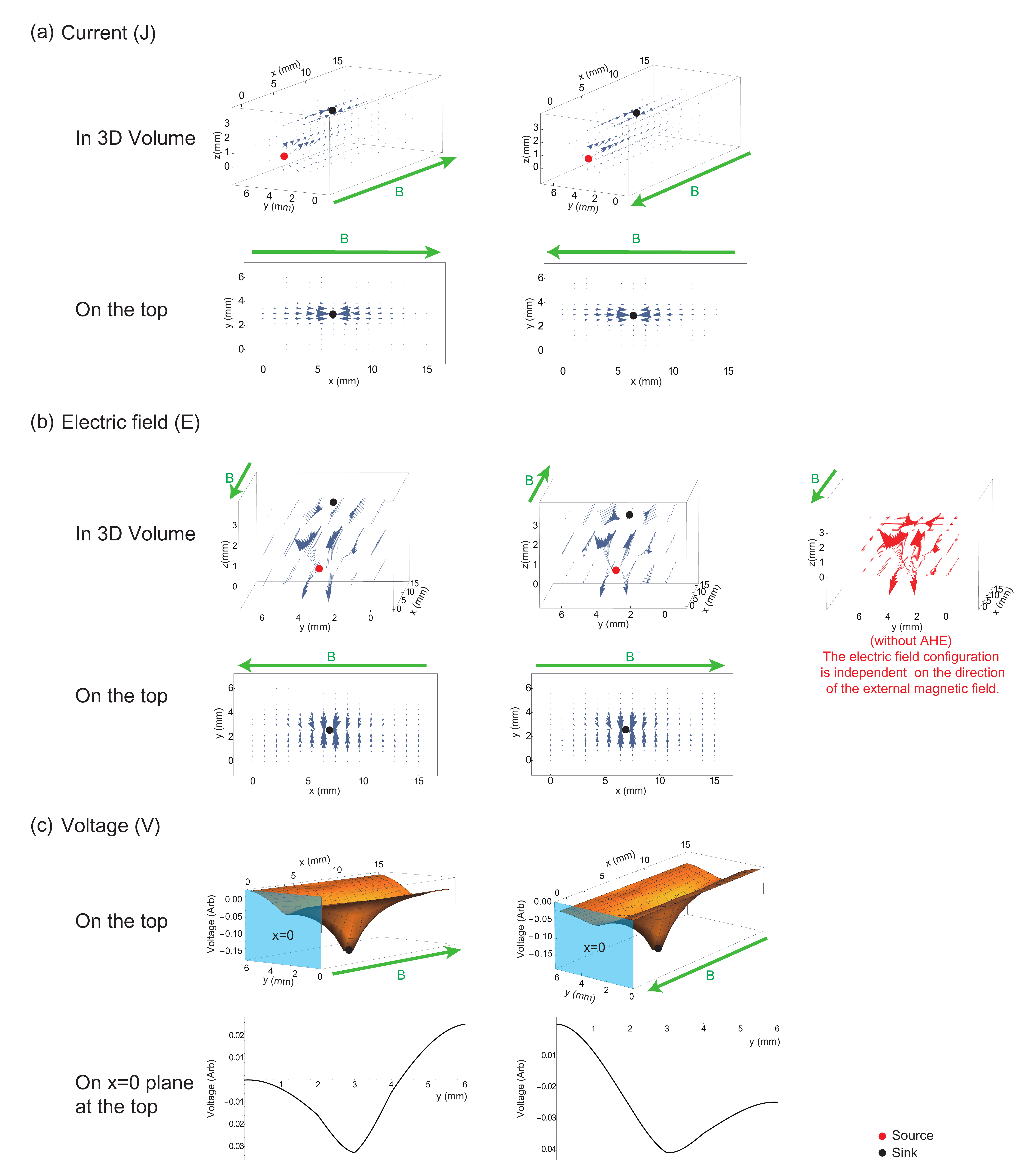}
\caption{Axion electrodynamics simulation in the presence of both NLMR and TMR effects and the AHE when $\v{B} \parallel \hat{x}$ and $\v{B} \parallel -\hat{x}$. (a) Current vector configuration in 3D volume and from the top. Nonlocal back flows still exist in the presence of the AHE. (b) Electric field configuration in 3D volume and from the top. A similar result without the AHE is shown as the 3D plot with red arrows for comparison. The electric field vector at each position is rotated in a clock-wise (counter clock-wise) fashion on the $\hat{x}$ axis when $\v{B_{ext}} \parallel \hat{x}$ ($\v{B_{ext}} \parallel - \hat{x}$) compared to the result without the AHE (simulation result with red arrows). This magnetic field dependent rotation does not occur in the absence of the AHE (in the presence of the NLMR). (c) The voltage drop on the top of the sample with an additional linear slope along the $y$ direction, compared to the result without the AHE. The voltage drop on the $x = 0$ plane seen from the top is shown for the clear presentation. The sign of the linear slope changes, depending on the direction of the magnetic field. As a result, the AHE is also generated over the entire sample in a nonlocal fashion along the transverse direction of the external magnetic field.}
\label{3daxionBxtotal}
\end{figure*}

Now, we introduce not only both NLMR and TMR but also AHE into the simulation of the axion electrodynamics. Let us consider the longitudinal case ($\v{B_{ext}} \parallel \v{J_{ext}}$) first. A current pattern of this case is similar to that of the previous case considering magnetoconductivity only. However, the electric field and voltage drop pattern are different because of the AHE effect. The electric field vector at each position is rotated in a clock-wise (counter clock-wise) fashion on the $\hat{x}$ axis when $\v{B_{ext}} \parallel \hat{x}$ ($\v{B_{ext}} \parallel - \hat{x}$). See Fig. \ref{3daxionBxtotal} (b). This rotation does not occur in the absence of the AHE (in the presence of the NLMR). Furthermore, the voltage drop on the top of the sample has a linear slope along the $y$ direction and its sign is changing depending on the direction of the magnetic field. See the voltage pattern in Fig. \ref{3daxionBxtotal} (c). This leads the voltage difference to depend on $y$ as a result of the AHE effect. Here, one important point is that when the nonlocal voltage drop is generated along the longitudinal direction of the external magnetic field, the AHE is also generated over the entire sample in a nonlocal fashion along the transverse direction of the external magnetic field.

Now, let us consider whether the nonlocal voltage drop is possible or not in the transverse case ($\v{B_{ext}} \perp \v{J_{ext}}$). Both NLMR and TMR effects are dominant in the previous case while the AHE effect ($2 \alpha g \v{E} \times \v{B}$) plays a central role in this case. For simplicity, let us focus on the first order of $2 \alpha g$ in Eq. (\ref{selfconsistentE}) and set $\v{B} = \v{B_{ext}}$. It turns out that this approximation gives a reasonably good answer close to the ``exact" result, where the external magnetic field itself is important.

When the external magnetic field is applied along the $z$ axis, the current-electric field equations read
\begin{eqnarray}
\sigma_x E_x &\approx& J'_x + 2 \alpha g (- B_{ext} \frac{J'_y}{\sigma_y}) , \nonumber \\
\sigma_y E_y &\approx& J'_y + 2 \alpha g ( B_{ext} \frac{J'_x}{\sigma_x}) , \nonumber \\
\sigma_z E_z &\approx& J'_z .
\end{eqnarray}
Ohm's law works in the $z$ direction. On the other hand, the field and current equations in the $xy$ direction can be expressed in a matrix form as follows
\begin{equation}
\left(\begin{array}{cc} \sigma_x & 0 \\
0 & \sigma_y \end{array}\right) \left(\begin{array}{c} E_x \\ E_y \end{array}\right) =
\left(\begin{array}{cc} 1 & -\frac{\sigma_{Weyl}}{\sigma_y} \\
\frac{\sigma_{Weyl}}{\sigma_x} & 1 \end{array}\right) \left(\begin{array}{c} J'_x \\ J'_y \end{array}\right) ,
\end{equation}
where $\sigma_{Weyl} = 2 \alpha g B_{ext}$. If we set $\sigma_x \sim \sigma_y \equiv \sigma_t$, the matrix equation is given by
\begin{equation}
\left(\begin{array}{c} E_x \\ E_y \end{array}\right) =
\frac{1}{\sigma_t}\left(\begin{array}{cc} 1 & -\frac{\sigma_{Weyl}}{\sigma_t} \\
\frac{\sigma_{Weyl}}{\sigma_t} & 1 \end{array}\right) \left(\begin{array}{c} J'_x \\ J'_y \end{array}\right)
\end{equation}
or
\begin{equation}
\left(\begin{array}{c} J'_x \\ J'_y \end{array}\right) =
\frac{1}{1+(\sigma_{Weyl}/\sigma_t)^2}\left(\begin{array}{cc} \sigma_t & \sigma_{Weyl} \\
-\sigma_{Weyl} & \sigma_t \end{array}\right) \left(\begin{array}{c} E_x \\ E_y \end{array}\right) . \label{anomalousHall}
\end{equation}
Here, $\sigma_t$ represents the transverse magnetoconductivity.

When the strength of the external magnetic field in the $z$ direction is weak, the transverse conductivity is not sufficiently reduced and the Hall effect itself is also negligible. Then, the asymmetry of the conductivity is not strong enough to generate any nonlocal effects. On the other hand, with strong enough magnetic fields, the AHE conductivity ($\sigma_{Weyl}$) is enhanced linearly proportional to $B$, but the transverse magnetoconductivity ($\sigma_t$) is reduced in the order of $1/B$. Recall that the external magnetic field is set to make $\sigma_{Weyl}:\sigma_{t}$ as $20:1$ in this simulation.

\begin{figure*}
\centering
\includegraphics[width=16cm]{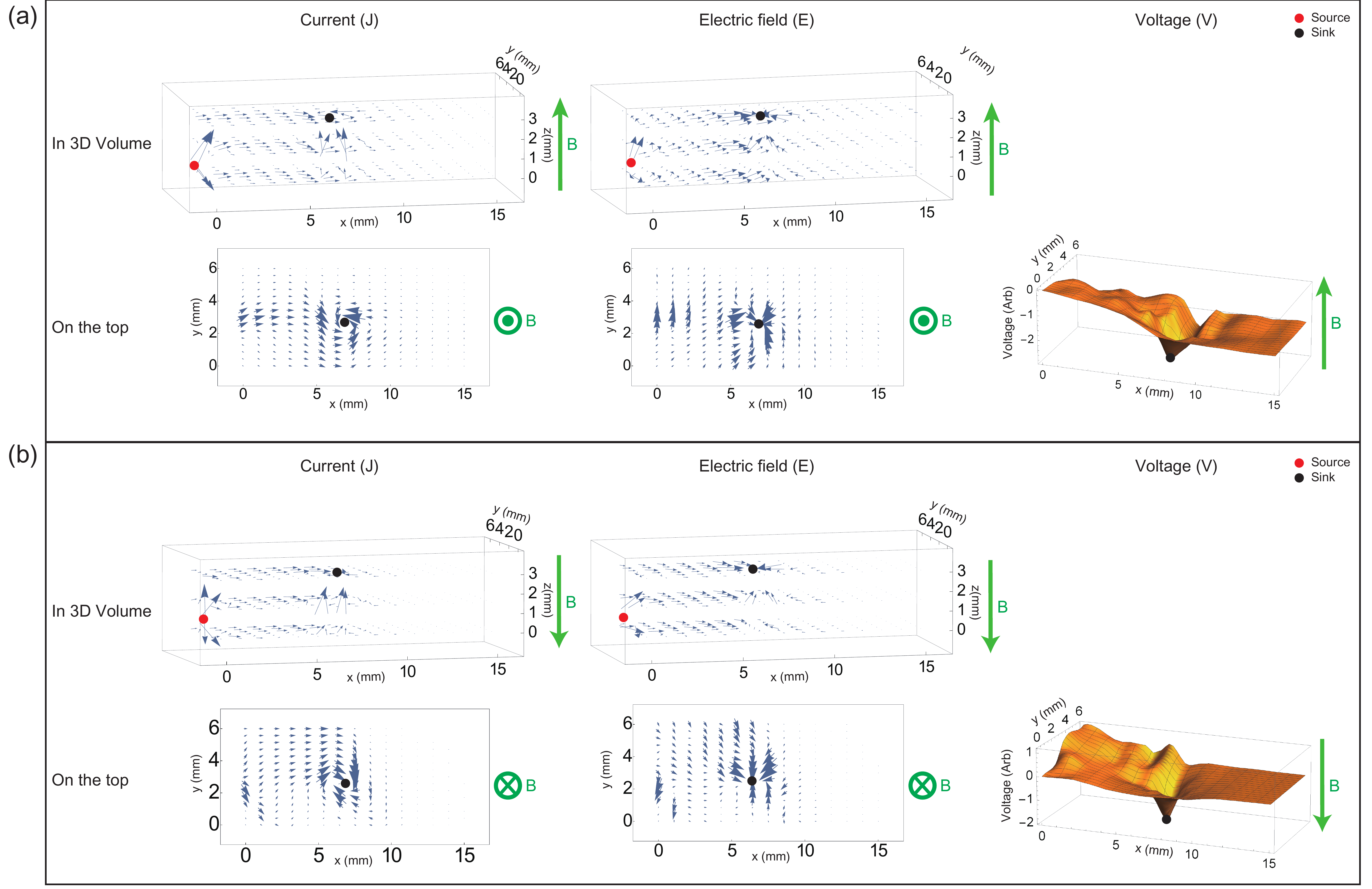}
\caption{Axion electrodynamics simulation in the presence of both NLMR and TMR effects and the AHE when (a) $\v{B} \parallel \hat{z}$ and (b) $\v{B} \parallel -\hat{z}$. When the anomalous Hall conductivity is much larger than the transverse magnetoconductivity ($\sigma_{Weyl} \gg \sigma_{t}$), a vortex-like current pattern is induced. The electric field configuration shows a significantly nonlocal pattern. See the text for more details.}
\label{3daxionBztotal}
\end{figure*}

When the anomalous Hall conductivity is much larger than the transverse magnetoconductivity ($\sigma_{Weyl} \gg \sigma_{t}$), a vortex-like current pattern is induced. The electric field configuration shows a significantly nonlocal pattern. Fig. \ref{3daxionBztotal} shows this situation. The current along the $z$ direction near the sink is mostly absorbed by the sink in an almost direct way. However, the current in the $x$ or $y$ direction near the sink is not directly absorbed by the sink but shows a vortex-like pattern due to the following reasoning. First, we are considering a static case. $\curl \v{E}$ vanishes identically and the corresponding electric-field configuration can not have a rotating pattern. Second, all the electric field lines near the sink have to go into the sink. Then, the current near the sink should rotate because the current on the $xy$ plane is almost perpendicular to the electric field in the $\sigma_{Weyl} \gg \sigma_{t}$ case. As seen near the core of the vortex pattern in Fig. \ref{3daxionBztotal}, the electric field is going into the core point. The voltage pattern with the transverse magnetic field ($\v{B_{ext}} \perp \v{J_{ext}}$) becomes nonlocal in this case. Furthermore, there exists asymmetry with the sign of the magnetic field. See the current pattern in Fig. \ref{3daxionBztotal} (a) and (b), where the rotating direction of each vortex configuration depends on the sign of the magnetic field. In principle, these effects could also occur with the conventional Hall effect only if the Hall effect is sufficiently large. But, we point out that the Hall conductivity can dominate in Weyl metals due to both AHE and TMR.

\section{Conclusion}

In this study, we examined macroscopic nonlocal voltage phenomena under external magnetic fields in a TRSB Weyl metal state.
%
%
%
When the external current is applied in parallel with the external magnetic field, the nonlocal voltage drop is generated with back flow currents over the almost entire region of the sample. In addition, the AHE is measured in the perpendicular direction to the external magnetic field. On the other hand, the nonlocal voltage drop can be generated around the sink with a vortex current pattern when the external current is applied in perpendicular with the external magnetic field. This nonlocal transport phenomena results from both the NLMR and TMR with the AHE in the axion electrodynamics.

\begin{acknowledgments}
K.-S. Kim was supported by the Ministry of Education, Science, and Technology (NRF-2021R1A2C1006453 and NRF-2021R1A4A3029839) of the National Research Foundation of Korea (NRF). We thank Jeehoon Kim and Dongwoo Shin for helpful discussions on their experiments.
\end{acknowledgments}

\section*{Appendix}
\setcounter{figure}{0}
\setcounter{equation}{0}
\setcounter{section}{0}
\renewcommand{\theequation}{A. \arabic{equation}}
\renewcommand{\thefigure}{A.\arabic{figure}}

\subsection{Justification of the Poynting theorem in the axion electrodynamics \label{Poynting}}

In this appendix, we show that the Poynting theorem is valid in the axion electrodynamics. Applying divergence to Eq. (\ref{Maxwell4}), we obtain
\begin{eqnarray}
\div{(\curl{\v{B}})} &=&  0 \nn
\rightarrow \div{(\mu \v{J} + \mu \epsilon \pd{\v{E}}{t} - 2\alpha g \mu \v{{\v{B}}} \times \v{E})} &=& 0 .
\end{eqnarray}

A derivative with respect to time for the electric field can be reformulated with the Poynting vector ($\v{S} = \v{E} \times \v{H}$) in the following way
\begin{eqnarray}
\pd{\div{\epsilon \v{E}}}{t} &=& -\div{\v{J}} + 2 \alpha g \div{(\v{B} \times \v{E})} \nn
&=& -\div{\v{J}} - 2\alpha g \mu(\div{\v{S}}) . \label{Poyntingvector}
\end{eqnarray}

Resorting to the vector identity ($\div{(\v{A} \times \v{B})}=\v{B} \cdot (\curl{\v{A}}) - \v{A}\cdot \curl{\v{B}}$), we rewrite the first line in Eq. (\ref{Poyntingvector}) as
\begin{eqnarray}
\pd{\div{\epsilon \v{E}}}{t} &=& -\div{\v{J}} +2\alpha g (\v{E} \cdot (\curl{\v{B}}) - \v{B} \cdot (\curl{\v{E}}))\nn
&=& -\div{\v{J}} +2\alpha g (\v{E} \cdot (\mu \epsilon \pd{\v{E}}{t} + \mu \v{J} - 2\alpha g \mu \v{B} \times \v{E}) \nn
&& + \v{B} \cdot \dot{\v{B}}) \nn
&=& -\div{\v{J}} +2\alpha g (\mu \epsilon \v{E} \cdot \dot{\v{E}} +\mu \v{J} \cdot \v{E} + \v{B} \cdot \dot{\v{B}}) \nn
&=& -\div{\v{J}} +2\alpha g \mu( \partial_t(\epsilon \v{E}^2 + \v{B}^2/\mu)/2  + \v{J} \cdot \v{E}) \nn
&=& -\div{\v{J}} +2\alpha g \mu( \partial_t u_{em} + \v{J} \cdot \v{E}) . \label{Poyntingthm}
\end{eqnarray}
Note that the third and fourth Maxwell equations (Eqs. (\ref{Maxwell3}) and (\ref{Maxwell4})) are incorporated in the first line of Eq. (\ref{Poyntingthm}). The anomalous Hall effect term ($2\alpha g \mu \v{B} \times \v{E}$) from $\curl{\v{B}}$ disappears in the third line of Eq. (\ref{Poyntingthm}) because $\v{E} \cdot (\v{B} \times \v{E})$ in the second line has to vanish. Comparing Eq. (\ref{Poyntingvector}) with Eq. (\ref{Poyntingthm}), one can immediately notice that the Poynting theorem $-\pd{u_{em}}{t} = \div{\v{S}} + \v{J} \cdot \v{E}$ is still valid in the axion electrodynamics.

\subsection{2D simulation with conventional Maxwell equations \label{2Dcase}}

\setcounter{figure}{0}
\setcounter{equation}{0}
\renewcommand{\theequation}{B. \arabic{equation}}
\renewcommand{\thefigure}{B.\arabic{figure}}

\begin{figure*}
\centering
\includegraphics[width=16cm]{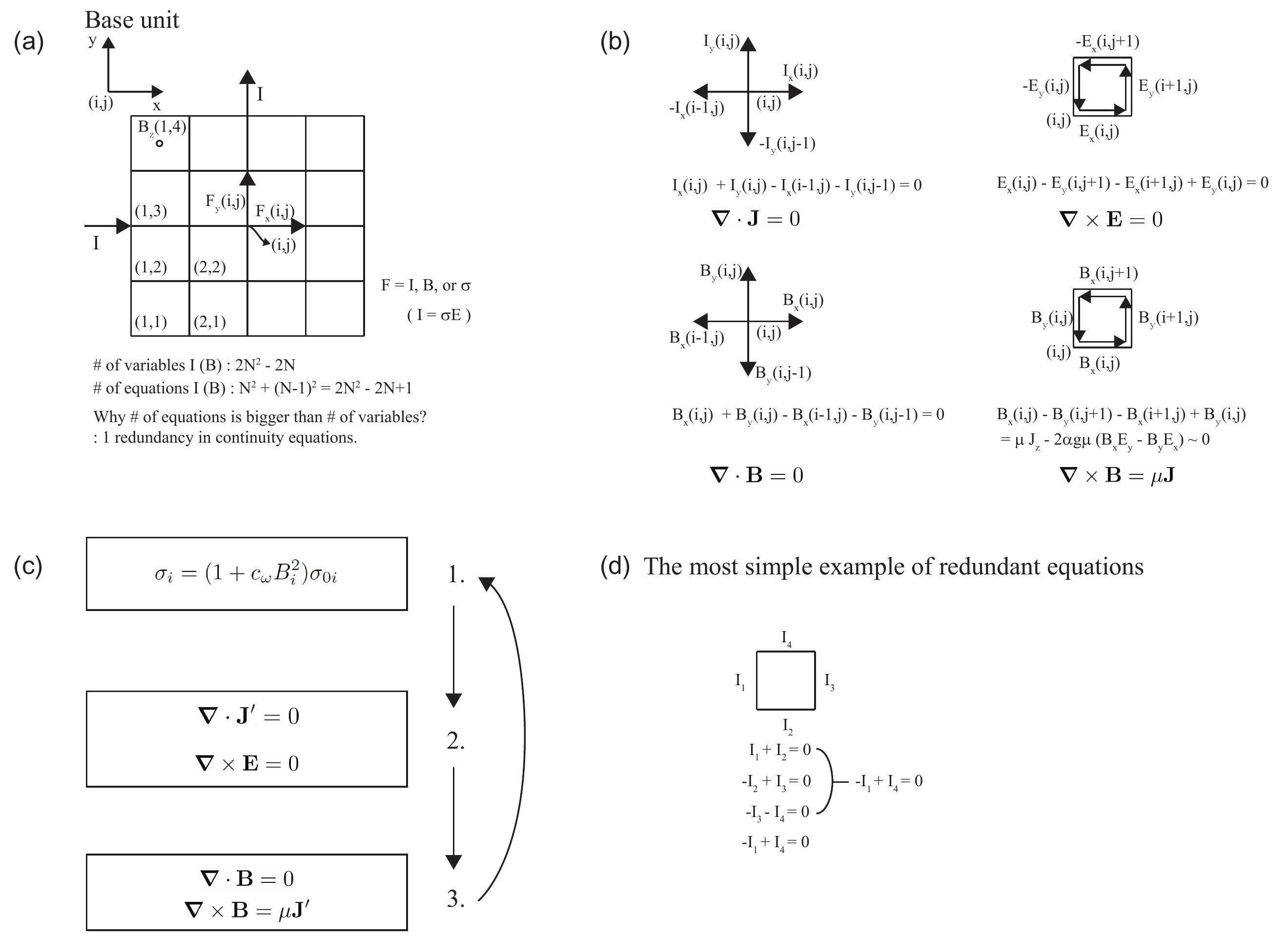}
\caption{How to simulate the Maxwell electrodynamics in 2D grid. (a) 2D grid structure with variables $F$ ($F = I$, $B$, or $\sigma$ at each point) with unit vectors $\hat{x}$ and $\hat{y}$. Number of variables, equations, and constraints are shown. See the text for more details. (b) Conversion of divergence and curl equations into algebraic equations on the grid structure. Divergence is converted as sum of four vectors at every vertex point, whereas curl is converted as sum of four vectors at every face of the unit square of the grid. (c) Simulation protocol for currents $I$, magnetic fields $B$, and conductivity $\sigma$. In the first step, we evaluate the conductivity first with an ansatz $\v{B}=\v{B_{ext}}$ at every point. Note that the conductivity would depend on the magnetic field if there are magnetoresistivity effects. Here, we set it to be a constant at every point because we assume no magnetoresistance effects ($\sigma(i,j) = \sigma_0$) for simplicity. In the second step, we obtain the solution of currents $\v{J'}$ at every point from divergence and curl equations of the current $\v{J'}$ consistent with the boundary current (i.e., $\v{J'}=\v{J'_{ext}}$ at the source point and sink point). Constituent equation between $\v{E}$ and $\v{J}$ should be used (which is nothing but the Ohm's law) to get the electric field $\v{E}$ from the current $\v{J'}$. Incorporating the solution of the conserved current $\v{J}'$ into the third step, the magnetic field $\v{B}$ can be evaluated from its divergence and curl equations. (d) A simple 2D grid example (square lattice with 4 points) shows one more redundancy in divergence equations. There are 4 equations from 4 points in curl equations, but summing over three equations immediately gives the other one.}
\label{2dsim}
\end{figure*}

In this appendix, we show a simple example for the conventional Maxwell equation in a 2D grid, presenting how to simulate the electric field and current. We consider a grid structure with a cartesian coordinate. Each point $(i,j)$ is defined by the $x$ component $i$ and the $y$ component $j$, where $i$ and $j$ are natural numbers. Each point has three variables ($\v{J}$, $\v{B}$, and $\boldsymbol{\sigma}$), and they are represented as arrows linking adjacent points with two directions ($x$ and $y$). See Fig. \ref{2dsim} (a).

Based on this grid construction, we solve four Maxwell equations with a current and charge source. The divergence or curl in Maxwell equations can be converted into summation over points or unit surface in an algebraic form on the grid structure as shown in Fig. \ref{2dsim} (b). Converting all differential equations at every point into algebraic equations on the grid, we can evaluate the conductivity and electric/magnetic field numerically from the three steps explained in Fig. \ref{2dsim} (c).

In this simple example of the 2D case, we consider $\v{B} = \v{B_{ext}}$ which is the simplest case for dominant magnetic fields. This corresponds to a situation of considering the Ohm's law without any magnetoresistance effects or spatially dependent $\theta$ terms. We recall that magnetoresistance effects are considered in the first step for the 3D Weyl metal case. The first and third Maxwell equations are contained in the second step. Actually, divergence equations ($\div{\v{J'}}=0$) correspond to continuity equations, and curl equations ($\curl \v{E} = 0$) correspond to the Kirchhoffs' law in a circuit theory if we convert $\v{J}'$ into the electric field $\v{E}$ using the Ohm's Law. In the third step, the second and fourth Maxwell equations are contained, but we don't have to calculate this step for this simple example because the magnetic field is fixed as an external magnetic field. All we have to do is to calculate $\v{J}'$ in the second step.

Before showing the result, we discuss how to count the number of redundancies in the divergence and curl equations.
%
%
Every vertex has two variables of $I_x$ and $I_y$ in this situation. Here, we resort to the Ohm's law, as discussed before.
%
%
It seems that we have $2N^2$ variables. However, $I_x$ should be zero at the rightmost edge, and $I_y$ should be zero at the top edge. Therefore, 2$N$ variables are already determined, and we have actually $2N^2 - 2N$ variables. Every point in the grid gives one divergence equation (continuity equation). Then we get $N^2$ equations if the grid is $N \times N$. On the other hand, every square gives one curl equation so $(N-1)^2$ equations should be considered additionally. Here we have a problem because we have one more number of equations than the number of variables. This indicates that one more redundancy exists in the equations and it is in the continuity equations for the 2D case. This redundancy problem can be eliminated by summing over all the continuity equations except one point when all divergence terms ($\div \v{E}$ and $\div \v{B}$) are zero all over the point. The result of the summation will give exactly the same equation of one missing point. A simple example of the redundancy problem is presented in Fig. \ref{2dsim} (d). Eliminating one redundant equation, we have the same number of equations and variables, and thus, we can find the solution of these $2N^2-2N$ variables by solving $2N^2-2N$ coupled linear equations.

%
%
%
%
%

\begin{figure*}
\centering
\includegraphics[width=16cm]{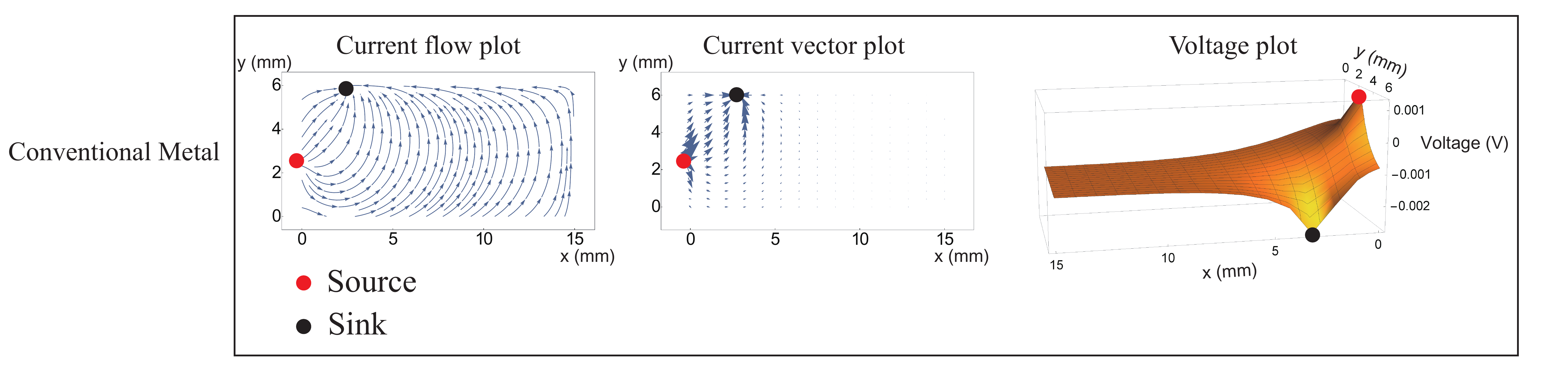}
\caption{Current $\v{J}$ and voltage V simulation in a 2D grid structure for conventional Maxwell equations with Ohm's law.}
\label{IEV2dtotal}
\end{figure*}

The 2D simulation result for conventional Maxwell equations is shown in Fig. \ref{IEV2dtotal}. There are no nonlocal electric field effects as expected.


\begin{thebibliography}{9}

\bibitem{WM1} F. D. M. Haldane, Phys. Rev. Lett. {\bf 93}, 206602 (2004).
\bibitem{WM2} S. Murakami, New J. Phys. {\bf 9}, 356 (2007).
\bibitem{WM3} A. A. Burkov and L. Balents, Phys. Rev. Lett. {\bf 107}, 127205 (2011).
\bibitem{WM4} P. Hosur, Phys. Rev. B {\bf 86}, 195102 (2012).


\bibitem{CME1} K. Fukushima, Dmitri E. Kharzeev, and Harmen J. Warringa, Phys. Rev. D {\bf 78}, 074033 (2008).
\bibitem{CME2} K. Landsteiner, E. Megias, and F. Pena-Benitez, Phys. Rev. Lett. {\bf 107}, 021601 (2011).
\bibitem{CME3} D. T. Son and N. Yamamoto, Phys. Rev. Lett. {\bf 109}, 181602 (2012).
\bibitem{CME4} M. A. Stephanov and Y. Yin, Phys. Rev. Lett. {\bf 109}, 162001 (2012).
\bibitem{CME5} G. Basar, Dmitri E. Kharzeev, and H.-U Yee, Phys. Rev. B {\bf 89}, 035142 (2014).
\bibitem{CME6} J.-Y. Chen, D. T. Son, M. A. Stephanov, Ho-Ung Yee, and Yi Yin, Phys. Rev. Lett. {\bf 113}, 182302 (2014).
\bibitem{CME7} C. Manuel and Juan M. Torres-Rincon, Phys. Rev. D {\bf 90}, 076007 (2014).
\bibitem{Boltzmann_Chiral_Anomaly1} J.-Wei. Chen, S. Pu, Q. Wang, and X.-N. Wang, Phys. Rev. Lett. {\bf 110}, 262301 (2013).
\bibitem{Boltzmann_Chiral_Anomaly2} D. T. Son and B. Z. Spivak, Phys. Rev. B \textbf{88}, 104412 (2013).
\bibitem{Boltzmann_Chiral_Anomaly3} Y.-S Jho and K.-S. Kim, Phys. Rev. B {\bf 87}, 205133 (2013).
\bibitem{Boltzmann_Chiral_Anomaly4} K.-S. Kim, H.-J. Kim, and M. Sasaki, Phys. Rev. B \textbf{89}, 195137 (2014).
\bibitem{Boltzmann_Chiral_Anomaly5} K.-S. Kim, Phys. Rev. B \textbf{90}, 121108(R) (2014).
\bibitem{Boltzmann_Chiral_Anomaly6} G. Sharma, P. Goswami, and S. Tewari, Phys. Rev. B \textbf{93}, 035116 (2016).
\bibitem{Boltzmann_Chiral_Anomaly7} I. Jang, J.-H. Han, and K.-S. Kim, Phys. Rev. B \textbf{95}, 054117 (2017).
\bibitem{Boltzmann_Chiral_Anomaly8} Yong-Soo Jho, Jae-Ho Han, and Ki-Seok Kim, Phys. Rev. B \textbf{95}, 205113 (2017).
\bibitem{Boltzmann_Chiral_Anomaly9} K.-M. Kim, D. Shin, M. Sasaki, H.-J. Kim, J. Kim, and K.-S. Kim, Phys. Rev. B {\bf 94}, 085128 (2016).
\bibitem{AHE1} A. A. Zyuzin and A. A. Burkov, Phys. Rev. B \textbf{86}, 115133 (2012).
\bibitem{AHE2} P. Goswami and Sumanta Tewari, Phys. Rev. B \textbf{88}, 245107 (2013).
\bibitem{AHE3} Y. Chen, D. L. Bergman, and A. A. Burkov, Phys. Rev. B \textbf{88}, 125110 (2013).
\bibitem{AHE4} Iksu Jang and Ki-Seok Kim, Phys. Rev. B {\bf 97}, 165201 (2018).

\bibitem{Axion_EM} F. Wilczek, Phys. Rev. Lett. {\bf 58}, 1799 (1987).

\bibitem{Dynamical_Axion_Review} Akihiko Sekine and Kentaro Nomura, J. Appl. Phys. {\bf 129}, 141101 (2021).

\bibitem{Axion_EM_Exp_TI_I} V. Dziom, A. Shuvaev, A. Pimenov, G. V. Astakhov, C. Ames, K. Bendias, J. Bottcher, G. Tkachov, E. M. Hankiewicz, C. Brune, H Buhmann, and L. W. Molenkamp, Nat. Commun. \textbf{8}, 15197 (2017).
\bibitem{Axion_EM_Exp_TI_II} L. Wu, M. Salehi, N. Koirala, J. Moon, S. Oh, and N. P. Armitage, Science \textbf{354}, 1124 (2016).
\bibitem{Axion_EM_Exp_TI_III} M. Li, W. Cui, L. Wu, Q. Meng, Y. Zhu, Y. Zhang, W. Liu, and Z. Ren, Can. J. Phys. \textbf{10}, 1139 (2014).
\bibitem{AxionEMreferee1} R.-Y. Zhang, Y.-W. Zhai, S.-R. Lin, Q. Zhao, W. Wen and M.-L. Ge, Sci. Rep. \textbf{5}, 13673 (2015).
\bibitem{AxionEMreferee2} A. A. Zyuzin and V. A. Zyuzin, Phys. Rev. B \textbf{92}, 115310 (2015).
\bibitem{AxionEMreferee3} Z. Qiu, G. Cao and X.-G. Huang, Phys. Rev. D \textbf{95}, 036002 (2017).
\bibitem{Axion_EM_Th_WM} Mehdi Kargarian, Mohit Randeria, and Nandini Trivedi, Sci. Rep. \textbf{5}, 12683 (2015).
\bibitem{AxionEMreferee4} S. Zhong, J. Orenstein and J. E. Moore, Phys. Rev. Lett. \textbf{115}, 117403 (2015).

\bibitem{nonlinear1} L. Wu, S. Patankar, T. Morimoto, N. L. Nair, E. Thewalt, A. Little, J. G. Analytis, J. E. Moore, and J. Orenstein, Nat. Phys. {\bf 13}, 350 (2017).
\bibitem{nonlinear2} J. Ma, Q. Gu, Y. Liu, J. Lai, P. Yu, X. Zhuo, Z. Liu, J.-H. Chen, J. Feng, D. Sun, Nat. Mater. {\bf 18}, 476 (2019).
\bibitem{nonlinear3} H. Rostami and M. Polini, Phys. Rev. B {\bf 97}, 195151 (2018).
\bibitem{nonlinear4} T. Morimoto, S. Zhong, J. Orenstein, and J. E. Moore, Phys. Rev. B {\bf 94}, 245121 (2016).

\bibitem{Axionprb} J.-H. Yang, J.-H. Kim, and K.-S. Kim, Phys. Rev. B {\bf 98}, 075203 (2018).

\bibitem{nonlocaldiffuse} S. A. Parameswaran, T. Grover, D. A. Abanin, D. A. pesin, and A. Vishwanath, Phys. Rev. X {\bf 4}, 031035 (2014).
\bibitem{nonlocalPW} Zhe Hou and Qing-Feng Sun, Phys. Rev. Res. {\bf 2}, 023236 (2020).
\bibitem{nonlocaleeinter} B. rosenstein, H. C. Kao, and M. Lewkowicz, Phys. Rev. B {\bf 95} 085148 (2017)
\bibitem{nonlocalhydro1} E. V. Gorbar, V. A. Miransky, I. A. Shovkovy, and P.O. Sukhachov, Phys. Rev. B {\bf 97}, 121105 (2018).
\bibitem{nonlocalhydro2} E. V. Gorbar, V. A. Miransky, I. A. Shovkovy, and P.O. Sukhachov, Phys. Rev. B {\bf 98}, 035121 (2018).

\bibitem{TSB_WM1} Heon-Jung Kim, Ki-Seok Kim, J.-F. Wang, M. Sasaki, N. Satoh, A. Ohnishi, M. Kitaura, M. Yang, and L. Li, Phys. Rev. Lett. \textbf{111}, 246603 (2013).
\bibitem{TSB_WM2} Dongwoo Shin, Yongwoo Lee, M. Sasaki, Yoon Hee Jeong, Franziska Weickert, Jon B. Betts, Heon-Jung Kim, Ki-Seok Kim, and Jeehoon Kim, Nat. Mater. {\bf 16}, 1096-1099 (2017).
\bibitem{ISB_WM1} J. Xiong, S. K. Kushwaha, T. Liang, J. W. Krizan, M. Hirschberger, W. Wang, R. J. Cava, and N. P. Ong, Science \textbf{350}, 413 (2015).
\bibitem{ISB_WM2} H. Li, H. He, H.-Z. Lu, H. Zhang, H. Liu, R. Ma,	Z. Fan,	S.-Q. Shen, and J. Wang, Nat. Commun. \textbf{7}, 10301 (2015).
\bibitem{ISB_WM3} X. Huang, L. Zhao, Y. Long, P. Wang, D. Chen, Z. Yang, H. Liang, M. Xue, H. Weng, Z. Fang, X. Dai, and G. Chen, Phys. Rev. X \textbf{5}, 031023 (2015).
\bibitem{ISB_WM4} Su-Yang Xu et al. Nat. Phys. {\bf 11}, 748 (2015).
\bibitem{ISB_WM5} S.-Y. Xu et al. Science {\bf 349}, 613 (2015).
\bibitem{ISB_WM6} L. Yang et al. Nat. Phys. {\bf 11}, 728 (2015).
\bibitem{ISB_WM7} Z. K. Liu, Nat. Mater. {\bf 15}, 27 (2016).


\bibitem{TMR} S.-B. Zhang, H.-Z. Lu, and S. Q. Shen, New. J. Phys. {\bf 18}, 053039 (2016).


\bibitem{EM_Textbook} John David Jackson, \textit{Classical Electrodynamics}, 3rd ed. (John Wiley $\&$ Sons., New York, 1999).

\end{thebibliography}
\end{document}